\documentclass[a4paper,twocolumn,english,pra,superscriptaddress]{revtex4}
\usepackage[T1]{fontenc}
\usepackage[latin1]{inputenc}
\usepackage{psfrag}

\makeatletter

\usepackage{graphicx} \usepackage{amsmath} \usepackage{amssymb}
\usepackage{babel} \makeatother

\begin{document}

\title{Oscillating Casimir force between impurities in one-dimensional Fermi liquids}

\author{J.N.~Fuchs}
\affiliation{Laboratoire de Physique des Solides, Universit\'e
Paris-Sud, CNRS, UMR 8502, F-91405 Orsay, France.}

\author{A.~Recati}
\affiliation{CRS BEC-INFM, Povo and ECT$^\star$, Villazzano, I-38050
Trento, Italy.}

\author{W.~Zwerger}
\affiliation{Technische Universit\"at M\"unchen, James Franck
Strasse, D-85748 Garching, Germany.}

\begin{abstract}
We study the interaction of two localized impurities in a repulsive
one-dimensional Fermi liquid via bosonization. In a previous paper
[Phys. Rev. A \textbf{72}, 023616 (2005)], it was shown that at
distances much larger than the interparticle spacing the impurities
interact through a Casimir-type force mediated by the zero sound
phonons of the underlying quantum liquid. Here we extend these
results and show that the strength and sign of this Casimir
interaction depend sensitively on the impurities separation. These
oscillations in the Casimir interaction have the same period as
Friedel oscillations. Their maxima correspond to tunneling
resonances tuned by the impurities separation.
\end{abstract}

\date{\today}
\maketitle

Interacting one-dimensional fermions are nowadays available in many
experimental systems. They can be electrons in semiconductor quantum
wires, in carbon nanotubes, in conducting polymers, etc.
\cite{Giamarchi} or fermionic atoms in ultra-cold gases \cite{MSKE}.
Impurities have a profound effect on the physical properties of
these low-dimensional systems. In certain exceptionally clean
systems, impurities can be created and controlled. For example, it
was recently proposed that atomic quantum dots -- i.e. individually
trapped atoms playing the role of impurities -- could be created in
an ultra-cold atomic gas \cite{RFZDZ}. Another possibility rests in
the ``buckles'' artificially created with AFM tips in single-wall
carbon nanotubes and which behave as tunnel junctions \cite{Dekker}.
As a last example, we mention atom wires that were recently created
on a Si(553)-Au surface and which featured vacancy-like defects that
were manipulated with an STM tip \cite{Snijders}. Once it is
recognized that impurities can be manipulated, one can ask new
questions in which they appear as main actors. For example, what is
the interaction between impurities which is mediated by the quantum
liquid of fermions? This problem was recently addressed in
\cite{RFPZ}. It was found that when the impurities are far apart
they behave like acoustic mirrors and interact via a Casimir-like
force. This is the expected behavior for any system whose low energy
properties are described in terms of phonons. However, for the
special case of non-interacting fermions, the calculation can be
done exactly and it was shown that this interaction features
oscillations as a function of the impurities separation \cite{RFPZ}.
In the present paper -- part II -- which can be considered as a
continuation of \cite{RFPZ} -- to which we shall refer as part I --
we extend some of these results to interacting fermions. Most
significantly, we show that the oscillation phenomenon in the
interaction remains essentially unaffected in the case of
interacting (repulsive) fermions. In the light of proofs showing the
absence of repulsive Casimir interactions for the photonic field
\cite{KK}, this is a quite remarkable situation. Here we have a
Casimir force whose strength and sign can be tuned by the impurities
separation.
%Another
%interesting question concerns resonant tunneling in double
%impurities structure. Resonant tunneling allows particles to pass
%easily through two impurities -- via a quasi-stationary state --
%even when each impurity is separately almost impenetrable. For
%symmetric impurities reaching a resonance requires tuning a single
%parameter \cite{KF}, which can be a gate voltage or the impurity
%separation. For example, by tuning a gate voltage it was possible to
%observe tunneling resonances in the conductance of a single-wall
%carbon nanotube interrupted by two artificially made buckles
%\cite{Dekker}. In the present paper, we extend our previous work
%\cite{RFPZ} by studying the influence of quasi-stationary states on
%the interaction between two impurities. In particular, we show how
%the Casimir force oscillates when tuning the inter-impurity
%distance.

In order to investigate that problem, we consider the following
model: two localized delta impurities embedded in a repulsive
one-dimensional Fermi gas, whose low-energy behavior is that of a
single channel Luttinger liquid with parameter $K\leq 1$. The
interaction between the impurities -- which is mediated by the
quantum fluctuations in the liquid -- is considered in the adiabatic
limit in which changing the distance between the impurities is slow
compared to the time the system takes to equilibrate. Within such a
limit, the interaction energy directly follows from the groundstate
energy. We should stress that we focus on interactions mediated by
quantum zero-point fluctuations and not on classical effects due to
modification of the average particle density. The latter dominate at
short distances of the order of the inter-particle distance and lead
to a kind of ``mattress effect''. For a discussion of the classical
ground state energy, see part I.

We first consider non-interacting fermions and discuss the
oscillations in the interaction energy in that case. Two delta
impurities located at $x_1=0$ and $x_2=r$ are embedded in a one
dimensional (1D) Fermi gas with Fermi velocity $v_F=\pi \rho_0/m$
where $\rho_0=k_F/\pi$ is the average density. The impurities
interact with the particles through a potential $g_\alpha
\delta(x-x_\alpha)$ where $\alpha=\{1,2\}$ and $g_\alpha$ are
microscopic scattering amplitudes. When the distance between the
impurities $r$ is much larger than the average inter-particle
distance $1/\rho_0$, it was shown in part I that the renormalized
interaction energy between the two impurities is given by
\begin{equation}
V_{12}=\frac{v_{F}}{2\pi r}\, \Re\, \mathrm{Li}_2
\big[-\frac{\gamma_1\gamma_2}{\sqrt{1+\gamma_1^2+\gamma_2^2+(\gamma_1\gamma_2)^2}}
e^{i\chi} \big]\, . \label{eq:Dilog}
\end{equation}
In the preceding equation, $\mathrm{Li}_2$ is the di-logarithmic
function, $\Re$ denotes the real part, $\gamma_\alpha=g_\alpha/v_F$
is the dimensionless impurity strength and we introduced the phase
$\chi(r,\gamma_\alpha)$ such that $\chi\equiv 2k_Fr+\delta$ modulo
$2\pi$ and $-\pi<\chi\leq \pi$. The phase shift
$\delta(\gamma_\alpha)=-\arctan \gamma_1-\arctan \gamma_2=-\arctan
[(\gamma_1+\gamma_2)/(1-\gamma_1\gamma_2)]$ reflects the time delay
of fermions scattering on delta impurities. As the function
$V_{12}/(v_F/r)$ is $2\pi$-periodic in $\chi$, the interaction
energy $V_{12}$ oscillates as a function of $r$. Maxima are
separated by $\Delta(2k_F r)=2\pi$, which corresponds to the average
inter-particle distance $\Delta r=1/\rho_0$, just as Friedel
oscillations. Maxima occur when $\cos(\chi/2)=\cos
(k_Fr+\delta/2)=0$. This is precisely the condition for tunneling
resonances in a non-interacting 1D Fermi gas, as shown in Appendix
\ref{tunres}. In Appendix \ref{disappear}, we discuss an interesting
situation in which the oscillations can disappear.

\begin{figure}[ptb]
\psfrag{r}{$2k_F r$} \psfrag{v}{$V_{12}/v_F k_F$}
\includegraphics[width=8cm]{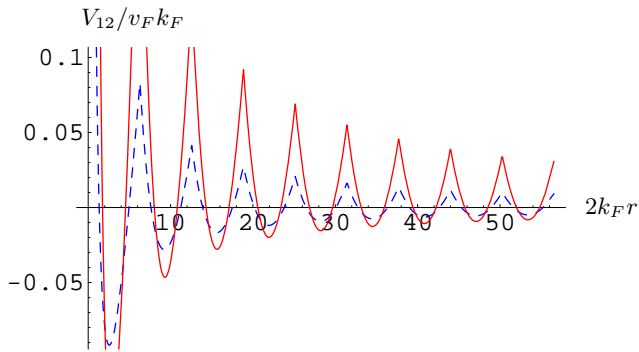}
\caption{Interaction energy $V_{12}$ between two identical strong
impurities separated by a distance $r$. Dashed line: non-interacting
fermions $V_{12}=v_F\Re\, \mathrm{Li}_2 (-e^{i \chi})/2\pi r$. Full
line: interacting fermions (with Luttinger parameter $K=0.6$)
$V_{12}=-\pi v_F/24Kr + v_F\chi^2/8\pi r K^2$. In both cases, $\chi=
2k_Fr+\delta$. For non-interacting fermions and strong impurities
($\gamma \gg 1$) the phase shift $\delta=-2\arctan \gamma \approx
-\pi$. For interacting fermions ($K<1$) and strong impurities
($\widetilde{\gamma} \gg 1$), we expect the phase shift to be
$\delta=-2\arctan (K^2 \widetilde{\gamma} )\approx -\pi$, see the
main text. The full line corresponds to this guess $\delta\approx
-\pi$.} \label{restun}
\end{figure}

In the limit of strong impurities ($\gamma_\alpha\gg 1$), the phase
shift $\delta \approx -\pi$ and Eq.~(\ref{eq:Dilog}) becomes
\begin{equation}
V_{12}\approx \frac{v_F}{2\pi r}\Re\, \mathrm{Li}_2 (-e^{i
\chi})=-\frac{\pi v_F}{24r}+\frac{v_F \chi^2}{8\pi r}\, ,
\label{eq:V12idealres}
\end{equation}
where $\chi \approx 2k_F r -\pi$. The last equation applies to
$-\pi<\chi\leq \pi$ with a periodic extension beyond that. The first
term has exactly the form of a 1D Casimir energy $-\pi v_F/24r$,
with the Fermi velocity $v_F$ playing the role of light velocity,
see e.g. \cite{Zee}. The second term introduces oscillations.
Indeed, the function $V_{12}/(v_F/r)$ is $2\pi$-periodic in $\chi$
and features wide maxima made of pieces of parabolas and centered at
$\chi=\pi$, see Figure \ref{restun}.
On the one hand, at a minimum, where $\chi=0$, the interaction
energy is given by the Casimir energy $-\pi v_F/24r$. This
correspond to a situation of no resonant tunneling
(anti-resonances). The Casimir energy has the usual interpretation
in terms of phononic ``radiation pressure'' on the impurities: the
outer radiation pressure of phonons overcomes the inner one and
hence the resulting attraction between the impurities. On the other
hand, at a maximum, where $\chi=\pi$, the interaction energy changes
sign and becomes positive reaching $\pi v_F/12r$. A physical
understanding of this change of sign in the Casimir force might be
obtained by noting that in a situation of resonant tunneling
($\chi=\pi$), fermions approaching the double barrier structure from
outside can tunnel with no reflection. There is thus no contribution
to the radiation pressure due to the phonons in the region outside
the impurities. By contrast, fermions which come from in between the
two barriers, have to tunnel through a single impurity and do not
benefit from resonant tunneling. At low energies, they are almost
perfectly reflected and therefore contribute a positive radiation
pressure on the impurities. At resonance therefore, the inner
radiation pressure is larger than the outer one, and thus the
effective interaction between the impurities is repulsive. In a
sense, this is like having impurities with transmission properties
that are not right-left symmetric and makes a connection to Imry's
idea of tuning the sign of the QED Casimir effect using asymmetric
mirrors \cite{Imry}.

We now come to the main point of part II. We show that the basic
oscillation phenomenon contained in Eq~(\ref{eq:V12idealres}), and
found in part I for non-interacting fermions, remains essentially
intact when including interactions between the fermions. In order to
study 1D interacting fermions, we use the Luttinger liquid formalism
\cite{Haldane} and concentrate on repulsive interactions. For
simplicity, we consider spinless fermions, for which the low-energy
description is given by the following hydrodynamic action
\begin{equation}
S_0=\frac{1}{2\pi K}\int \!\!\mathrm{d} x\,\int_0^\beta
\!\!\mathrm{d}\tau \left[ u
(\partial_{x}\theta)^{2}+\frac{1}{u}(\partial_{\tau}\theta)^{2}
\right]\, , \label{eq:LLaction}
\end{equation}
where $\beta=1/T$ is the inverse temperature and $\theta$ is related to the
density of the liquid by
\begin{equation}
\rho(x)=\left(\rho_{0}+ \partial_x \theta/\pi\right) \left[1+
2\cos(2\theta+ 2 k_{F} x)+\dots \right], \label{density}
\end{equation}
where $\rho_{0}$ is the equilibrium density and only the first
harmonics are written \cite{Haldane}. The Luttinger liquid
description depends on three parameters: the sound velocity $u$, the
Luttinger interaction parameter $K$ and a cutoff energy $\omega_c
\sim u k_F$ above which this effective description breaks down.
Repulsive short-range interactions imply $0<K<1$. With long-range
(Coulomb) interactions, the system becomes a 1D Wigner crystal,
which can be seen as the $K\to 0$ limit of the Luttinger liquid
\cite{GRS,Schulz}, but can not be treated within the present
formalism. Non-interacting fermions correspond to $K=1$ and $u=v_F$.
If there is full translational invariance $uK=v_F$ \cite{Haldane}.

We consider, as before, two localized impurities sitting at position $x_1=0$ and
$x_2=r$. The interaction between the impurities and the Luttinger
liquid is
\begin{equation}
S_{i}=\int_{0}^{\beta}\mathrm{d}\tau\!\sum_{\alpha=1,2}
\widetilde{g}_\alpha \rho(x_{\alpha})\, , \label{eq:Sidens}
\end{equation}
where $\widetilde{g}_\alpha$ is a phenomenological scattering
amplitude. As the Luttinger liquid is an effective description valid
at low energy, the phenomenological scattering amplitude
$\widetilde{g}_\alpha$ is not equal to the microscopic scattering
amplitude $g_\alpha$. The precise relation between the two
scattering amplitudes is not generally known. In the renormalization
group language, the phenomenological scattering amplitude
$\widetilde{g}$ is roughly a running scattering amplitude at the
energy scale $\omega_c$. For simplicity, we consider the symmetric
case $\widetilde{g}_1=\widetilde{g}_2=\widetilde{g}$. Inserting the
density expansion (\ref{density}) into the interaction
(\ref{eq:Sidens}) gives rise to several different terms. The first
term $2\beta \widetilde{g}\rho_0$ is just twice the classical energy
cost of adding an impurity to the liquid: it does not contribute to
the renormalized interaction energy $V_{12}$. The second term due to
forward scattering is proportional to $\partial_{x}\theta$. This
term is often eliminated via a gauge transformation: $\theta(x) \to
\theta(x)-K\widetilde{g}(\mathcal{H}(x-r)+\mathcal{H}(x)-1/2)/u$,
where $\mathcal{H}(x)$ is the Heaviside step function. However, this
operation is not totally innocent here and forward scattering can
not be eliminated: the gauge transformation gives rise to a phase
shift $\delta=-2K\widetilde{g}/u$ in the $2k_F$ backscattering term.
In the limit of weak impurities, this phase shift agrees with the
one found for non-interacting fermions $\delta=-2\arctan(g/v_F)
\approx -2g/v_F$. Actually, we expect that the phase shift in the
interacting case is given by $\delta=-2\arctan(K\widetilde{g}/u)$
for \emph{arbitrary} impurity strengths, however to our knowledge
there are no exact results on this problem. In fact, the precise
value of $\delta$ depends on the detailed potential and the behavior
of the fermion liquid at high energies where a LL description no
longer applies. This issue is often discussed in the context of the
x-ray edge singularity or in the Kondo problem and is related to the
order in which two non-commuting limits are taken: sending the size
of the impurity to zero versus sending the bandwidth to infinity.
When using the Luttinger liquid formalism, one automatically takes
the second limit first, see Ref.~\cite{GNT}. The third term due to
$2k_F$ backward scattering gives the dominant contribution to the
action $S_i$ and, when including the phase shift $\delta$, appears
as
\begin{equation}
S_{i}\approx V \int_0^{\beta}\mathrm{d}\tau
(\cos[2\theta(0)]+\cos[2\theta(r)+\chi])\ , \label{cosine}
\end{equation}
where $V=2\widetilde{g}\rho_0$ and $\chi$ is a phase such that $\chi
\equiv 2k_Fr + \delta$ modulo $2\pi$ and $-\pi<\chi \leq \pi$. There
are also higher order terms coming from the density harmonics,
however their naive scaling dimension shows that they are less
relevant and we therefore neglect them. Actually perturbative
renormalization group calculations show that $2k_F$ backscattering
is always relevant as long as $K<1$ while all higher backscattering
operators $\cos(2n\theta(x)+2nk_Fx)$, with $n=2,3,\dots$, are
irrelevant as long as $K>1/n^2$ \cite{KF,FN}. As we neglect those
higher backscattering operators, our model only describes the
situation of two impurities in an interacting Fermi gas such that
$1/4<K<1$.

As a side remark, we note that there are other ways of coupling
impurities to a bosonic bath of oscillators, such as a Luttinger
liquid. For example, in the context of QED Casimir force studies
\cite{Jaffe}, the impurities (mirrors) are usually taken to couple
directly to the density of the photonic field, i.e. $\theta^2$
rather than $\cos 2\theta$ would appear in equation~(\ref{cosine}).
This is the case in particular for the theorem of Kenneth and Klich
on the absence of a repulsive Casimir force \cite{KK}. Another
example was considered in Ref.~\cite{RFZDZ}, where only forward
scattering was kept, resulting in a mapping to the spin-boson model.
These different models, which all describe two impurities in a bath
of quantum oscillators, do not necessarily show the same behavior.

From now on, we shall work with the action $S=S_0+S_i$ with $S_i$
given by Eq.~(\ref{cosine}). The partition function is $Z=\int
\mathcal{D}\theta(x,\tau) \exp(-S)$ and the integral on the field
$\theta(x,\tau)$ can be performed everywhere except at the position
of the impurities $x=x_\alpha$. In this way we are left with an
action $A=A_0+A_i$ for the fields $\theta_\alpha(\tau)\equiv
\theta(x_\alpha,\tau)$ only
\begin{eqnarray}
A_0&=&\sum_{n \in \mathbb{Z}} \frac{f_n}{\pi K}(|\theta_{1,n}|^2+
|\theta_{2,n}|^2-e_n(\theta_{1,-n} \theta_{2,n} +\text{c.c.}))\nonumber \\
A_i&=&V \int_0^{\beta}\mathrm{d}\tau
(\cos[2\theta_1]+\cos[2\theta_2+\chi])\, ,
\end{eqnarray}
where $\omega_n$ are bosonic Matsubara frequencies, the Fourier
transform is defined as $\theta_n=T\int d\tau \exp(i\omega_n \tau)
\theta(\tau)$, the characteristic finite-size frequency is
$\omega_r= u/r$ and we defined $e_n= \exp(-|\omega_n|/\omega_r)$ and
$f_n= \beta |\omega_n|/(1-e_n^2)$. Up to an overall normalization
factor, the partition function is now given by $Z=\int
\mathcal{D}\theta_\alpha(\tau) \exp(-A)$. The partition function
gives the free energy $F=-T\ln Z$, from which the renormalized
interaction energy follows: $V_{12}(r)=F(r)-F(\infty)$. From the
parametric dependance of the action $A$ on the phase $\chi$, we see
that the free energy (and therefore $V_{12}$) will be
$2\pi$-periodic in $\chi$. As $\chi = 2k_Fr+\delta$, this means that
$V_{12}$ will have some structure with periodicity $1/\rho_0$ in $r$
just as for non-interacting fermions. Therefore, the interaction
energy $V_{12}$ oscillates as a function of the impurities
separation $r$.

We first evaluate the partition function using the saddle point
approximation. For small $K$ and strong impurities
$\widetilde{\gamma}=\widetilde{g}/v_F\gg 1$, the fields
$\theta_\alpha$ are pinned to the values that minimize $A_i$. We
therefore expand the cosines around a minimum to quadratic order in
the fields. It should be noted that terms independent of the fields
give an important contribution here \footnote{These terms, which
give rise to the oscillations in the interaction energy $V_{12}$,
were neglected in the simplified calculations of Ref.~\cite{RFPZ}.}.
The resulting free energy is:
\begin{eqnarray}
F_{s}(V)&=&F_{s}(0)+\frac{\omega_r \chi^2}{8\pi K}\frac{2\pi K
V}{\omega_r
+ 2\pi K V}  \\
&+&\frac{T}{2}\sum_n \ln (1+\frac{2\pi K
V}{|\omega_n|}(1+e_n))(1+\frac{2\pi K V}{|\omega_n|}(1-e_n))
\nonumber
\end{eqnarray}
As the validity of the saddle point approximation requires that
$2\pi K V \gg \omega_r$, the renormalized interaction becomes:
\begin{eqnarray}
V_{12}&=& \frac{\omega_r \chi^2}{8\pi K} \\
&+& \frac{T}{2}\sum_n \ln \frac{1+\frac{2\pi K
V}{|\omega_n|}(1+e_n)}{1+\frac{2\pi K
V}{|\omega_n|}}\frac{1+\frac{2\pi K
V}{|\omega_n|}(1-e_n)}{1+\frac{2\pi K V}{|\omega_n|}}
\nonumber
\end{eqnarray}
In the limit of zero temperature $\omega_r \gg T$, it gives
\begin{eqnarray}
V_{12}-\frac{\omega_r \chi^2}{8\pi K}&\approx&
\frac{\omega_r}{2\pi}\int_0^{\omega_c/\omega_r} dx \ln(1-e^{-2x}) \nonumber \\
&\approx &-\frac{\pi \omega_r}{24}\, , \label{eq:V12res}
\end{eqnarray}
where we used $\omega_c \gg \omega_r$ \footnote{This result can also
be written $V_{12}/\omega_r =\pi (1-K)/24K +\Re\, \mathrm{Li}_2
(-\exp i\chi) /2\pi K$.}. This result generalizes
Eq.~(\ref{eq:V12idealres}) when interactions between particles are
taken into account. It also admits an interpretation as an
attractive Casimir energy $-\pi\omega_r/24$ -- albeit with the true
sound velocity $u$ replacing the Fermi velocity $v_F$ -- modified by
oscillations contained in the term $\omega_r \chi^2/8\pi K$, see
Figure \ref{restun}. The width of the maxima measured relative to
their amplitude is proportional to $K$. Therefore the maxima are
narrower in the presence of interactions between the particles. The
exact position of the maxima is given by the values of $r$ such that
$\chi=\pi$, i.e. $\cos(\chi/2)=\cos(k_Fr+\delta/2)=0$, and depends
on both the impurity strength and the inter-particle interactions
through the phase shift $\delta$. As in the case of non-interacting
fermions, this is exactly the condition at which tunneling
resonances occur in a Luttinger liquid, see the Appendix
\ref{tunres}.
%The line shape of the peaks in the interaction energy
%$V_{12}$ has a singularity in the first derivative and is different
%from the well-known conductance resonances \cite{KF,FN}. In
%particular, the maxima in the interaction energy $V_{12}$ have a
%finite width at $T=0$.

At this stage, it is worth mentioning that there is a mapping of our
problem -- as given by actions (\ref{eq:LLaction}) and
(\ref{cosine}) -- on the double-boundary sine-Gordon model studied
in the context of conformal field theory \cite{CSS}. For special
values of the Luttinger parameter $K=1/[2(1+n)]$, where $n$ is a
strictly positive integer (the so-called reflectionless points), the
authors of Ref.~\cite{CSS} were able to calculate exactly the
groundstate energy of this model \footnote{Note that
$K=1/[2(1+n)]\leq 1/4$.}. In the limit of strong impurities and for
the reflectionless points, it yields the same interaction energy
$V_{12}=-\pi \omega_r/24+\omega_r \chi^2 (1+n)/4\pi$ as
Eq.~(\ref{eq:V12res}).

In order to study what happens for values of $K$ close to one, we
turn to the self-consistent harmonic approximation (SCHA), which is
just an implementation of Feynman's variational principle, see, e.g.,
\cite{Giamarchi}. Expanding the cosines in Eq.~(\ref{cosine}) to
second order in the fields, we replace the coefficient $2V$ in front
of the quadratic term by a variational parameter $\lambda$. The true
action $S=S_0+S_i$ is now replaced by a quadratic variational action
$S_v$. The parameter $\lambda$ is determined self-consistently by
finding a minimum to the variational free energy $F_v(\lambda)=-T\ln
Z_v +T\langle S-S_v \rangle$. The latter is given by
\begin{equation}
F_v(\lambda)=F_{s}(0)+f-\lambda \frac{\partial f}{\partial \lambda}
-2V\exp(-\frac{\partial f}{\partial \lambda})
\end{equation}
where
\begin{eqnarray}
f&=& F_{s}(V=\lambda/2)-F_{s}(0)\nonumber \\
&\approx& -\frac{\pi \omega_r}{24}+\frac{\omega_r \chi^2}{8\pi
K}+\frac{\omega_c}{\pi}\ln (1+\frac{\pi K \lambda}{\omega_c})
+K\lambda \ln (1+\frac{\omega_c}{\pi K \lambda})\nonumber
\end{eqnarray}
the second equality being justified when $\omega_c \gg \omega_r \gg
T$ and $\pi K\lambda \gg \omega_r$ which is a validity condition of
the SCHA. Extremizing the variational free energy gives:
\begin{equation}
\lambda=2V\exp(-\frac{\partial f}{\partial
\lambda})=2V\left(\frac{\pi K\lambda}{\pi
K\lambda+\omega_c}\right)^K
\end{equation}
When $K<1$, this equation has two solutions: $0$ (corresponding to a
maximum of $F_v$) and $\bar{\lambda}>0$ (corresponding to a
minimum), where $\bar{\lambda}\approx 2V$ when $2\pi V\gg \omega_c$
and $\bar{\lambda}\approx 2V(2\pi KV/\omega_c)^{K/(1-K)}$ when $2\pi
V \ll \omega_c$ \footnote{When $K=1$, the solution giving a minimal
free energy is $\bar{\lambda}=\max (0,2\pi V/\omega_c-1)$.}. As
neither $\bar{\lambda}$ nor $\partial f (\bar{\lambda})/\partial
\lambda$ depends on $r$, the renormalized interaction energy is
given by $V_{12}\approx
f(\bar{\lambda},r)-f(\bar{\lambda},\infty)\approx -\pi
\omega_r/24+\omega_r \chi^2/8\pi K$. The result obtained in the
saddle point approximation Eq.~(\ref{eq:V12res}) is recovered and is
therefore valid more generally for $K<1$ as long as $\pi
K\bar{\lambda} \gg \omega_r$, i.e. for strong impurities or large
distances between the impurities.

The situation of weak impurities or short distances (but always in
the limit $r\gg 1/\rho_0$) can be studied thanks to a perturbative
calculation of the partition function. In the regime $K>1/2$, the
interaction energy is
\begin{equation}
V_{12} = - \widetilde{\gamma}^2 \frac{v_F}{2\pi r}\cos \chi \times
\frac{(uk_F)^2
K\Gamma(K-1/2)}{\omega_c^{2K}\omega_r^{2-2K}\sqrt{\pi}\Gamma(K)}\ ,
\label{eq:genRKKY}
\end{equation}
where $\Gamma(x)$ is the gamma function and $\chi \approx 2k_Fr$
when $\widetilde{\gamma} \ll 1$, as already shown in part I. This is
a RKKY-type interaction, which decays with a renormalized power law
$r^{1-2K}$. It is valid as long as $|V_{12}|\ll \omega_r$, i.e.
$r\ll \rho_0^{-1}\widetilde{\gamma}^{1/(K-1)}$. In a similar way, we
obtain
\begin{equation}
V_{12} = -\widetilde{\gamma}^2 \cos \chi \times
\frac{2(v_Fk_F)^2\beta}{\pi^2(1-2K)(1-K)(\beta \omega_c)^{2K}}\, ,
\label{eq:ctcos}
\end{equation}
in the regime $0\leq K<1/2$ and
\begin{equation}
V_{12} = -\widetilde{\gamma}^2 \cos \chi \times
\frac{4(v_Fk_F)^2}{\pi^2 \omega_c}\ln (\beta \omega_r)\, ,
\end{equation}
for the special case $K=1/2$, again provided that $\omega_r \gg T$.
For $0\leq K<1/2$, the interaction given by Eq.~(\ref{eq:ctcos})
still shows Friedel oscillations but its envelope does not depend on
the inter-impurity distance $r$ anymore, which reflects the tendency
of the system toward Wigner crystallization when $K$ goes to zero.
However the validity of this calculation is restricted to small but
finite temperature as $|V_{12}|\ll \omega_r$ corresponds to $r\ll
u\omega_c^{2K}T^{1-2K}/(v_Fk_F\widetilde{\gamma})^2$. This means
that when $T\to 0$, the domain of validity of perturbation theory
shrinks to zero, when $0\leq K\leq 1/2$. We note that, at the
reflectionless point $K=1/4$, Eq.~(\ref{eq:ctcos}) has the same
structure as the exact $T=0$ result $V_{12} = -\widetilde{\gamma}^2
\cos \chi \times (v_F k_F)^2/(\pi \omega_c)$ \cite{CSS}, where $\chi
\approx 2k_Fr$ when $\widetilde{\gamma} \ll 1$. Again, as in the
case of strong impurities, maxima in the interaction energy $V_{12}$
occur when $\cos (\chi/2)=0$, which is the condition for tunneling
resonances in a Luttinger liquid, see the Appendix \ref{tunres}.

Tunneling resonances for the conductance are only found for
$1/4<K\leq 1$. Indeed on resonance $2k_F$ backscattering is absent
and higher order backscattering operators control the behavior of
the system. Once $K<1/4$, $4k_F$ backscattering, which we neglected,
becomes relevant and washes out the tunneling resonances \cite{KF}.
Therefore, in order to study what happens for very small values of
$K$, one should also take these other backscattering operators into
account when computing $V_{12}$. Using the methods of part I, the
present work could also be extended to treat spin $1/2$ fermions.

In conclusion, we found that at large impurities separation or for
strong impurities, the renormalized interaction between the
impurities results from a Casimir force modified by oscillations --
due to quasi-stationary states in between the impurities. The
oscillations can be spanned by changing the distance between the
impurities and are separated by the average inter-particle distance,
i.e. they have the same periodicity as Friedel oscillations. The
position of the maxima (peaks) is given by precisely the same
condition as for tunneling resonances and depends on particle
interactions -- through the Luttinger parameter $K$ -- and on the
impurity strength through a phase shift. At intermediate distances
and for Luttinger parameter $K>1/2$, the interaction is of the RKKY
type, featuring Friedel oscillations with an envelope which is
decaying as a power law with a renormalized exponent. When $1/4<K
<1/2$, Friedel oscillations are still present at finite $T$ but are
not decaying anymore. With ultracold fermionic atoms confined in a
cigar-shaped trap, as described in detail in part I, it should be
possible to detect this oscillating Casimir force.

\vspace{0.5cm}

\section*{Acknowledgments}
We acknowledge useful discussions with I.~Safi, H.~Saleur,
N.~Prokof'ev and P.~Faccioli. We also thank V.~Meden for very
instructive discussions and comments on the present manuscript.

\vspace{0.5cm}

\appendix
\section{Tunneling resonances}\label{tunres}
We first consider a single fermion at the Fermi surface -- with
incoming wave vector $k_F$ -- scattering on two delta impurities
$g_1\delta(x)$ and $g_2\delta(x-r)$. Using the transport matrix
formalism, it is easy to obtain its transmission probability across
the double barrier structure
\begin{eqnarray}
P(k_F)&=& \left[ 1+\gamma_1^2+\gamma_2^2+2\gamma_1^2\gamma_2^2 \right.  \\
&+& \left. 2\gamma_1 \gamma_2
\sqrt{(1+\gamma_1^2)(1+\gamma_2^2)}\cos(2k_Fr+\delta)\right]^{-1}\nonumber\,
,
\end{eqnarray}
where $\gamma_j=mg_j/k_F$ and $\delta=-\arctan \gamma_1 - \arctan
\gamma_2$. This probability has maxima -- called tunneling
resonances -- when $\cos (k_Fr+\delta/2)=0$. On resonance, the
energy of a quasi-stationary state in between the impurities matches
the Fermi energy. In the case of symmetric impurities
($\gamma_1=\gamma_2$), there is perfect transmission on resonance.
For asymmetric impurities, the transmission probability is always
smaller than one. The condition for tunneling resonances coincides
with the position of maxima of $V_{12}$.

We now turn to interacting fermions. Tunneling resonances in a
Luttinger liquid were first studied in Ref.~\cite{KF,FN} and are
reviewed by Giamarchi \cite{Giamarchi2}. In these works, the forward
scattering term on the impurities was not taken into account. When
this term is neglected, and in the absence of a gate voltage, the
criterion for tunneling resonances in a Luttinger liquid is
$\cos(k_Fr)=0$ \cite{KF,FN,Giamarchi2}. It is trivial to redo these
calculations starting from the action (\ref{cosine}) including the
phase shift $\delta=-2K\widetilde{g}/u$ resulting from forward
scattering on the impurities. The only modification amounts to
replace $2k_Fr$ by $2k_Fr+\delta$ and therefore the criterion for
tunneling resonances becomes $\cos(k_Fr+\delta/2)=0$. As in the case
of non-interacting fermions, this coincides with the position of the
maxima in $V_{12}$.

\section{Disappearance of the oscillations in $V_{12}$}
\label{disappear} Here we discuss an interesting phenomenon in the
interaction energy $V_{12}$, which was pointed out to us by V.~Meden
\cite{Meden}. Consider non-interacting spinless fermions on a
lattice in a tight-binding model with hopping amplitude $t$, lattice
spacing $b$, and filling factor $f=\rho_0 b=1/2$. The Fermi momentum
is $k_F=\pi f/b=\pi/(2b)$ and the Fermi velocity is $v_F=2tb\sin
(f\pi)=2tb$. This model is expected to have the same low energy
behavior as the continuum model considered in the present paper.
There are also two site impurities with dimensionless strength
$\gamma_1$ and $\gamma_2$ separated by a distance $r$, which can
only take discrete values $r=\text{integer}\times b$. Therefore
$2k_Fr=\pi/(2b)\times\text{integer}\times b=\text{integer}\times
\pi$, which means that $\exp i\chi = (-1)^{r/b}\exp i\delta$, where
$\delta=-\arctan [(\gamma_1 +\gamma_2)/(1-\gamma_1 \gamma_2)]$. If
in addition the impurities are fined tuned such that $\gamma_1
\gamma_2=1$, the phase shift $\delta=-\pi/2$ and
Eq.~(\ref{eq:Dilog}) becomes:
\begin{eqnarray}
V_{12}&=&\frac{v_{F}}{2\pi r}\, \Re\, \mathrm{Li}_2
\left[\frac{i(-1)^{r/b}}{\gamma_1+\gamma_2}\right]\nonumber \\
&=&\frac{v_{F}}{8\pi r}\mathrm{Li}_2 [-(\gamma_1+\gamma_2)^{-2}]
\sim -\frac{v_F}{r} \ .
\end{eqnarray}
This shows that the interaction appears to be purely attractive --
as $\mathrm{Li}_2 (-1/4)\leq\mathrm{Li}_2
(-(\gamma_1+\gamma_2)^{-2})<0$ -- and not oscillating, when probing
inter-impurity distances $r$ on the lattice only. This phenomenon
occurs when three conditions are fulfilled: the system is
half-filled ($k_F=\pi/(2b)$), the impurity separation only takes
discrete values on a lattice ($r=\text{integer}\times b$) and the
impurity strengths satisfy $\gamma_1\gamma_2=1$. For a numerical
study of the half-filled lattice model of spinless fermions with
nearest-neighbor interaction, and with two impurities, showing that
this phenomenon survives in the case of interacting fermions, see
Ref.~\cite{Meden}.

\end{document}